\documentclass[aps,pra,twocolumn,amsmath,amssymb,superscriptaddress,tightenlines]{revtex4}
\usepackage{graphicx}
\usepackage{epsfig}
\usepackage{dcolumn}
\usepackage{bm}
\usepackage{amssymb}
\usepackage{bm}
\usepackage{amsmath, bm}
\usepackage{upgreek}
\usepackage[colorlinks,citecolor=blue,linkcolor=blue,hyperindex]{hyperref}

\begin{document}

\title{Fermion liquids as quantum Hall liquids in phase space: \\ A unified approach for anomalies and responses}

\author{Jaychandran Padayasi}
\affiliation{Department of Physics and National High Magnetic Field Laboratory,
Florida State University, Tallahassee, Florida 32306, USA}

\author{Ken K. W. Ma}
\affiliation{Department of Physics, Northeastern University, Massachusetts 02115, USA}
\affiliation{Department of Physics and National High Magnetic Field Laboratory,
Florida State University, Tallahassee, Florida 32306, USA}

\author{Kun Yang}
\affiliation{Department of Physics and National High Magnetic Field Laboratory,
Florida State University, Tallahassee, Florida 32306, USA}
\date{\today}


\begin{abstract}
    The discovery of many strongly correlated metallic phases has inspired different routes to generalize or go beyond the celebrated Landau Fermi liquid theory. To this end, from universal consideration of symmetries and anomalies, Else, Thorngren and Senthil (ETS) have introduced a class of theories called ersatz Fermi liquids which possess a Fermi surface and satisfy a generalized Luttinger's theorem. In this work, we view all such fermion liquids obeying the Luttinger theorem as incompressible quantum Hall liquids in higher-dimensional phase space and use it as the starting point to derive their effective low-energy field theory. The noncommutativity of phase space motivates us to use the Seiberg-Witten map to derive the field theory in an ordinary (commutative) space and naturally leads to terms that correspond to the correct topological Chern-Simons action postulated by ETS in one, two, and three dimensions. Additionally, our approach also reproduces all the non-topological terms that characterize important contributions to the response, including the semiclassical equations of motion. Finally, our derivations of Chern-Simons terms from the Seiberg-Witten map also verify a longstanding conjecture in noncommutative field theory. 
\end{abstract}

\maketitle

\section{Introduction}

Metallic phases formed by interacting electrons are of central importance in condensed matter physics and materials science. The most familiar of them is the Landau Fermi liquid, which can be connected smoothly to the free Fermi gas where quasi-electron and hole excitations are well defined around the Fermi surface~\cite{Coleman-book,Yang-book}. Recently, many experiments have shed light on a variety of ``non-Fermi~liquid" phases where the effects of interactions are nonperturbative, and quasiparticle excitations are generally absent—distinct from the behavior observed in Fermi liquids~\cite{Lee-NF-review}. Nevertheless, many known examples of non-Fermi liquids support Fermi surfaces that satisfy the Luttinger theorem, like the one-dimensional(1D) Luttinger liquid~\cite{Voit-1995}. Due to the nonperturbative nature of interaction, we do not have a general way to study non-Fermi liquids, except in one dimension where the powerful method of bosonization is available~\cite{Stone-book, Tsvelik-book, Giamarchi-book, vonDelft-review}.

In this paper, we demonstrate that considerable insights into fermion liquids (including both Fermi and non-Fermi liquids in any dimensions) can be gained by viewing them as incompressible quantum Hall liquids in phase space, as long as the Luttinger theorem is obeyed. A quantum Hall liquid could not be further from a Fermi liquid at first glance: The former is incompressible with a gap for bulk excitations, whereas the latter is compressible and supports gapless quasiparticle excitations around its Fermi surface in the bulk. The former requires breaking of time-reversal symmetry to exist in two dimensions, while there is no such requirement for the latter. Here, we point out that they are very closely related in the following way: A $d$-dimensional Fermi liquid is an incompressible liquid in {\em phase} space, with the Fermi surface being part of its phase-space {\it boundary}. We argue that this phase space incompressible liquid is nothing but an integer quantum Hall liquid in $2d$ spatial dimensions, whose gapless boundary excitations correspond to the quasiparticle excitations around the Fermi surface of the Fermi liquid. The key connection is the noncommutativity of phase space. We further argue that this analogy can be extended to a class of non-Fermi liquid metallic states termed ersatz Fermi liquid (EFL) by Else, Thorngren, and Senthil (ETS)~\cite{ETS}. There are no sharp quasiparticles there but the Fermi surface remains well defined and satisfies the Luttinger theorem (or more precisely, the Luttinger sum rule)~\cite{Luttinger-Ward, Luttinger, AGD-book}. In our picture, the Luttinger theorem is essentially a trivial consequence of the incompressibility in phase space. 

It is important to emphasize that our approach here is complementary to existing literature on the subject~\cite{Qi2015, Else-PRL2021, Wang-PRB2021, Lu2023, Yuxuan2024}. Instead of motivating the phase-space Chern-Simons (CS) term on the basis of an anomaly-matching argument, our proposed effective \textit{noncommutative} field theory for the incompressible liquid in phase space enables us to derive the anticipated CS term that gives rise to the ETS anomaly~\cite{ETS}\footnote{A derivation for a very special case can be found in the very short note~\cite{Wang-arxiv}.}. More importantly, our derivation yields all other non-topological responses in a controlled and systematic way. Thus, our work complements and significantly extends recent work on non-Fermi liquids through symmetries and anomalies.

\subsection{Motivation and nontechnical discussion}
\label{sec:motivation}

\begin{figure} [htb]
\centering
\includegraphics[width=3.2in]{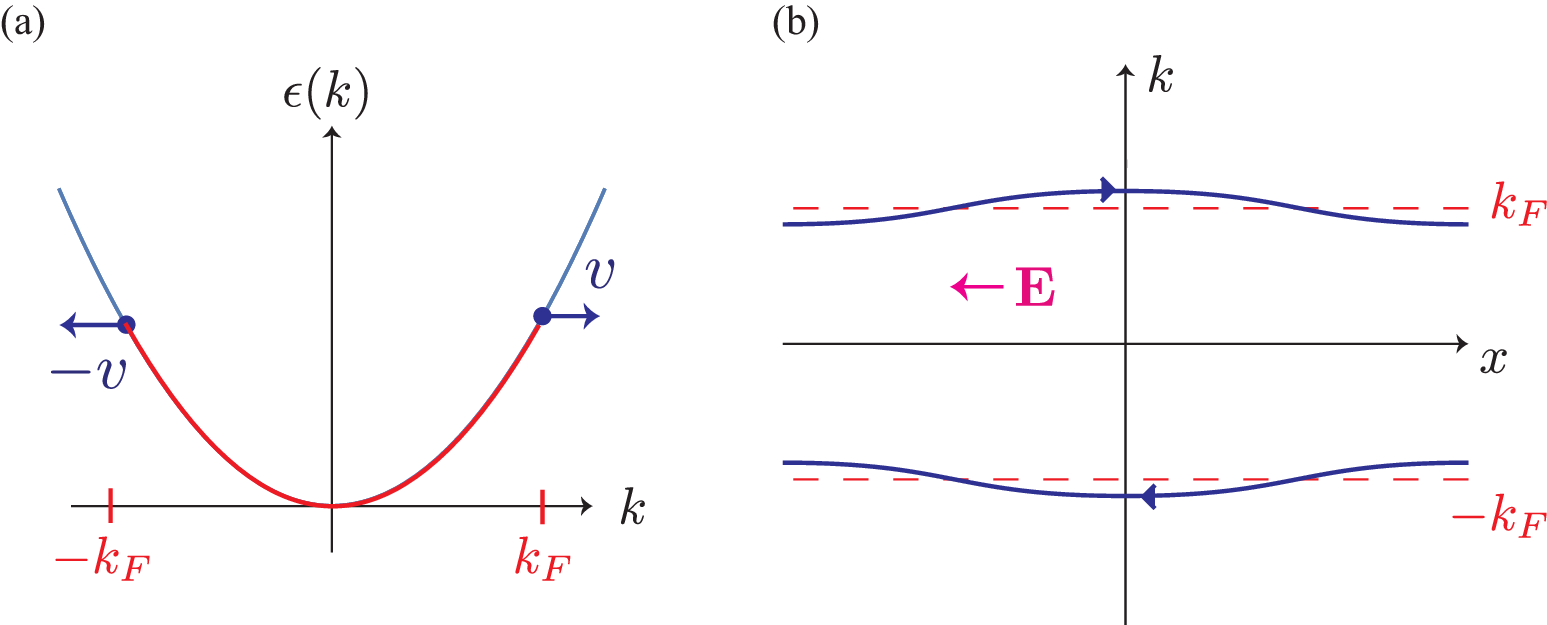}
\caption{One-dimensional Luttinger liquid (a) as an integer quantum Hall strip (b) in 2D. The strip is aligned along the horizontal (i.e., $x$) direction. The vertical direction is the linear momentum labeled by $k$, and the kinetic energy maps onto a confining potential that restricts the electrons to $-k_F < k < k_F$ (dashed lines) in the ground state. The low-energy excitations are the edge fluctuations (solid lines) that move to the left and right at the lower and upper edges, respectively. There is an anomaly when an electric field $\mathbf{E}$ (in the $x$ direction) is present, which pumps electrons from the lower edge to the upper edge via the bulk quantized Hall conductance in this analogy.}
\label{fig:QHstrip}
\end{figure}

Before diving into technicality, we first illustrate the ideas outlined above by considering the simplest case $d=1$ (1D), where the ersatz Fermi liquid is simply the familiar Luttinger liquid. We show that it can be viewed as an integer quantum Hall liquid in $d=2$ (2D), which is the dimension of the phase space of the Luttinger liquid. As shown in Ref.~\cite{Yang-PRA2020}, we can view the 2D phase space as the 2D real space subject to a strong perpendicular magnetic field, where all particles are confined to the lowest Landau level. In this representation, the two components of the guiding center, 
${\bf R} = (R_x, R_y)$ represent $(x, p)$ of the original 1D system if we set $\hbar$ and the magnetic length $\ell_B$ to be one, as $[R_x, R_y] =i\ell^2_B$~\cite{Yang-book} and $[x, p] = i\hbar$. The kinetic energy $p^2 \sim R_y^2$ thus becomes a confining potential along the $y$ direction, and the lowest Landau level electrons form a strip of $\nu=1$ quantum Hall liquid with width $2k_F$ aligned along the $x$ direction. Note that the Fermi points $\pm k_F$ become the $y$ coordinates of the edge locations, and the low-energy excitations of the Luttinger liquid are nothing but the familiar (chiral) edge excitations of the quantum Hall liquid, with left and right movers localized in the lower and upper edges respectively. The above discussion is illustrated in Fig.~\ref{fig:QHstrip}. Naively, the left and right moving electrons are separately conserved corresponding to an emergent U(1)$\times$U(1) symmetry. However, the extra U(1) symmetry (in addition to overall charge conservation) suffers a 't Hooft anomaly, which, in the 2D formulation is nothing but the bulk-edge correspondence of the quantum Hall liquid: electrons move from the lower edge to the upper edge through the bulk when an electric field is applied in the $x$ direction due to the quantized Hall effect, ruining the separate conservation of the left and right movers. Just like for $2$D $\nu=1$ quantum Hall liquid, this anomaly is captured by the Chern-Simons action of the external gauge potential $A$ in the $2+1$-D {\em phase space} for Luttinger liquid:
\begin{equation}
S[A] 
=\frac{i}{4\pi}\int~A \wedge dA
=\frac{i}{4\pi}\int~d^3 x~\epsilon^{\alpha\beta\gamma}A_\alpha\partial_\beta A_\gamma
\label{eq:1D}
\end{equation}
which is the 1D version of the ETS anomaly~\cite{ETS}.

We now generalize the consideration above to $d>1$ spatial dimensions, whose phase-space dimension becomes $2d$, with $d$ pairs of conjugate variables,
\begin{equation}
[x_m, p_n]= i\delta_{mn}.
\label{eq:Commutator}
\end{equation}
A Fermi liquid or ersatz Fermi liquid that satisfies the Luttinger theorem is, by definition, an {\em incompressible} liquid in {\em phase space}. This is because the Luttinger theorem states that the momentum-space volume enclosed by the Fermi surface is~\cite{Luttinger}
\begin{equation}
\mathcal{V}_m=(2\pi)^d n_r,
\label{eq:Luttinger_Theorem}
\end{equation}
where $n_r$ is the real-space density of the liquid, implying the phase-space density of the liquid
\begin{equation}
n_r/\mathcal{V}_m=(2\pi)^{-d}
\end{equation}
is {\em fixed}, as long as we identify the Fermi surface as its boundary in momentum space, as we did in the 1D Luttinger liquid above. Similarly, the kinetic energy (which can correspond to some band dispersion, not necessarily quadratic) in the phase-space, that depends on the momentum coordinates only, maps onto a confinement potential in momentum space. As we will demonstrate, the phase-space action for such an incompressible liquid includes, among other terms, the topological CS term in $2d + 1$ spacetime dimensions:
\begin{equation}
S_{0}[A] = \frac{i}{(d+1)!(2\pi)^d}\int  A \wedge dA \cdots \wedge dA,
\label{eq:general D}
\end{equation}
where $\wedge dA$ is repeated $d$ times. This is precisely the action that gives rise to the ETS anomaly, which was asserted in Ref.~\cite{ETS} with no derivation.

At this point, let us briefly review the core result of Ref.~\cite{ETS} and related work. ETS identify that the microscopic translation symmetry in an ersatz Fermi liquid maps to a global symmetry $G_{\mathrm{IR}}$ of the infrared (IR) theory. In two spatial dimensions, the Fermi surface is parametrized by a single parameter $\theta$, and the appropriate symmetry is the \textit{loop group} LU(1)  of U(1). LU(1) is the set of all smooth maps from the circle $S^1$ to U(1), and reflects the fact that the quasiparticle density is conserved at each point on the Fermi surface. The kinematic properties of the ersatz Fermi liquids are governed by the `t Hooft anomaly of this emergent loop group symmetry, and the CS term in $2d+1$ space-time dimensions was postulated as the appropriate topological term describing the higher-dimensional symmetry-protected topological (SPT) phase whose boundary theory carries the 't Hooft anomaly for the $G_{\mathrm{IR}}$ gauge fields. The interpretation of the Chern-Simons action as a phase-space quantum Hall liquid was mentioned in passing in Ref.~\cite{ETS} and in earlier literature. Specifically, Ref.~\cite{Qi2015} analyzed the responses arising from the phase space Chern Simons term extensively. However, neither of these works take into consideration the noncommutativity of phase space. Using the analogy with (higher-dimensional) quantum Hall effect and the inherent noncommutativity of phase space, we are able to {\em derive} Eq.~\eqref{eq:general D}. This is the primary goal and main result in this work.

The higher-dimensional CS action has been discussed in literature in the context of higher-dimensional generalizations of the quantum Hall effect \cite{Zhang-Hu}. Earlier works by Karabali and Nair \cite{KN01, KN02, KN03} have argued that the $(2k+1)$-dimensional CS term describing the dynamics of a quantum Hall droplet on complex projective spaces $\textbf{CP}^k$ arises from a Seiberg-Witten type mapping. While the analogy to the higher-dimensional quantum Hall effect is important to our work as a motivation, our starting point is much simpler, that of an incompressible liquid in phase space. The key derivations rely heavily on the symplectic form for the noncommutative parameter that characterizes quantum phase space.

The organization of this manuscript is as follows. In Sec.~\ref{sec:review-NCFT}, we provide a brief review of noncommutative field theory and the Seiberg-Witten (SW) map \cite{Seiberg-Witten}. This serves as a comprehensive introduction to the key technique that we will use in formulating the field theory for the fermion liquid in noncommutative phase space. Readers who are familiar with noncommutative field theory may skip this section. In Sec.~\ref{sec:action}, we introduce the phase-space action in arbitrary dimensions, and discuss general considerations that hold for all the special cases discussed in later sections. In Sec.~\ref{sec:1D-FL}, we demonstrate the power of our approach by formulating a 1D fermion liquid (i.e., a Luttinger liquid) as the corresponding noncommutative field theory in two-dimensional phase space. By assuming that the fermion liquid satisfies Luttinger's theorem, we show explicitly how the CS term emerges naturally from the SW map. In this sense, we successfully derive the ETS anomaly in one-dimensional ersatz Fermi liquid~\cite{ETS}. Then, the same approach is employed for two-dimensional fermion liquids in Sec.~\ref{sec:2D-FL}. Here, we obtain the most important result: The CS term governing the anomaly and universal topological responses emerges from the second-order SW map. Along the way, we also obtain other terms that describe non-topological responses. These terms describe the usual electromagnetic responses in the semiclassical theory of metals, which are relevant and important in realistic systems. Next, we discuss fermion liquids in three dimensions (3D) and beyond in Sec.~\ref{sec:generalization}. Parallel to the 2D case, we obtain the seven-dimensional (7D) CS term from the third-order expansion of the SW map. Lastly, we summarize our work in Sec.~\ref{sec:conclusion}. The Appendixes discuss some technical issues and mathematical derivations using the SW map, which also form an important part of the present work.

\section{Review of noncommutative field theory} 
\label{sec:review-NCFT}

A noncommutative field theory is defined in a \textit{noncommutative} space with the coordinates satisfying~\cite{RMP-noncommutative, Szabo, Barbon, Wohlgenannt},
\begin{eqnarray} \label{eq:start}
\left[X^\mu, X^\nu\right]
=i\theta^{\mu\nu}.
\end{eqnarray} 
In this work, the noncommutative space is the quantum phase space with spatial coordinates $Q_j$ and momentum $P_j$, where $j=1,\cdots, d$. Here, $d$ is the spatial dimension of the system. By defining $X^{2j-1}=Q_j$ and $X^{2j}=P_j$, the noncommutative parameters take the form
\begin{eqnarray}
\label{eq:NCphaseSpace}
\theta^{2j-1~2j}=\hbar.
\end{eqnarray}
To easily compare the dimensionality between different terms in the later discussion, we will keep $\hbar=1$ explicit. 

A theory defined in noncommutative space $X^\mu$ can be traded for a noncommutative field theory defined on the usual commuting coordinates $x^\mu$ ($[x^\mu, x^\nu]=0$).
For this, the product between two noncommutative fields needs to be replaced by the Weyl-Moyal star product, which is defined as
\begin{eqnarray} \label{eq:MY-product}
f(\mathbf{x})\star g(\mathbf{x})
=
\exp\left(\frac{i}{2}\theta^{\mu\nu}
\frac{\partial}{\partial x^\mu}
\frac{\partial}{\partial y^\nu}
\right) 
\lim_{\mathbf{x}\rightarrow\mathbf{y}}
f(\mathbf{x}) g(\mathbf{y}).
\end{eqnarray} 
Our discussion in Sec.~\ref{sec:motivation} and early literature already suggested that noncommutative field theory has a deep connection with quantum Hall physics~\cite{Bellissard1994, Read1998, Susskind2001, Polychronakos2001, Hellerman2001, Fradkin2002, Pasquier2007, Cappelli2009, Wu2016}. This is further corroborated by the recent work on formulating a composite fermion theory restricted to the lowest Landau level and other related topics~\cite{Dong2020, Milovanovic2021, Du2021, Dong2022, Goldman2022, Ma-CF2022}. 

\subsection{Seiberg-Witten map}

The Seiberg-Witten map provides a systematic procedure to approximate a noncommutative field theory by a commutative field theory~\cite{Seiberg-Witten}. Here, we only focus on theories with a U(1) gauge field coupled to a matter field. In commutative field theories, a U(1) gauge potential transforms as
\begin{eqnarray} \label{eq:A-transform}
A_\mu \rightarrow A_\mu+\partial_\mu \lambda
~\Rightarrow~
\delta_\lambda A_\mu=\partial_\mu\lambda.
\end{eqnarray}
Both equations describe an exact transformation here. For a matter field transforming in the \textit{fundamental representation}, one has
\begin{eqnarray} \label{eq:psi-transform}
\psi \rightarrow e^{i\lambda}\psi
~\Rightarrow~
\delta_\lambda\psi=i\lambda\psi.
\end{eqnarray}
Note that the second equation describes an infinitesimal transformation. We will use this notation in the following discussion. The corresponding transformation for the noncommutative U(1) gauge potential needs to transform in the \textit{adjoint representation},
\begin{align} \label{eq:nC-A-transform}
\nonumber
&\hat{A}_\mu
\rightarrow \hat{A}_\mu
+\partial_\mu\Lambda
+i [\Lambda, \hat{A}_\mu]_\star
\\
\Rightarrow~
&\delta_\Lambda \hat{A}_\mu
=\partial_\mu\Lambda
+i [\Lambda, \hat{A}_\mu]_\star.
\end{align}
The last term in the transformation does not vanish since it is a star product, which marks the biggest difference between noncommutative and commutative gauge theories. For the noncommutative matter field, one has
\begin{align} \label{eq:nC-psi-transform}
\Psi
\rightarrow e^{i\Lambda}\star\Psi~
\approx
\Psi+i\Lambda\star \Psi
~\Rightarrow~
\delta_{\Lambda} \Psi
=i\Lambda\star \Psi.
\end{align}

It is emphasized that we only consider left-coupling gauge field in this work, which describes the probe field. This is different from the studies of composite fermion theory, in which both left- and right-coupling (i.e., the emergent) gauge fields are involved~\cite{Dong2020, Milovanovic2021, Du2021, Dong2022, Ma-CF2022, Goldman2022}. This avoids the subtlety of formulating a mutual noncommutative Chern-Simons term 
~\cite{Goldman2022, Du2021}. 

\subsubsection{Gauge equivalence conditions}

The Seiberg-Witten map relates the sets of noncommutative and commutative fields via a list of gauge equivalence conditions. From Eq.~\eqref{eq:psi-transform}, subsequent gauge transformations on $\psi$ commute. On the other hand, Eq.~\eqref{eq:nC-psi-transform} shows that subsequent U(1) gauge transformations for $\Psi$ do not commute. Specifically, one has
\begin{align}
\nonumber
&e^{i\Lambda_\alpha}\star e^{i\Lambda_\beta}\star \Psi
\approx\left(1+i\Lambda_\alpha\right)\star\left(1+i\Lambda_\beta\right)\star\Psi
\\
\Rightarrow~
& i\delta_\alpha\Lambda_\beta\star\Psi
=-(\Lambda_\alpha+\Lambda_\beta)\star\Psi-i\Lambda_\alpha\star\Lambda_\beta\star\Psi.
\end{align}
Reversing the order of the transformations, one finds
\begin{align}
\nonumber
&e^{i\Lambda_\beta}\star e^{i\Lambda_\alpha}\star \Psi
\approx\left(1+i\Lambda_\beta\right)\star\left(1+i\Lambda_\alpha\right)\star\Psi
\\
\Rightarrow~
&i\delta_\beta\Lambda_\alpha\star\Psi
=-(\Lambda_\alpha+\Lambda_\beta)\star\Psi-i\Lambda_\beta\star\Lambda_\alpha\star\Psi.
\end{align}
Since the above rules hold for every $\Psi$, we have
\begin{eqnarray} \label{eq:Lambda-eq}
i\delta_\alpha \Lambda_\beta
-i\delta_\beta \Lambda_\alpha
=-[\Lambda_\alpha, \Lambda_\beta]_\star.
\end{eqnarray}
This gives a consistency condition that the noncommutative gauge parameters $\Lambda_\alpha$ and $\Lambda_\beta$ need to satisfy.

Next, the gauge potentials $\hat{A}$ and $A$ are related by the following condition:
\begin{eqnarray} \label{eq:gauge-equivalence}
\hat{A}_\mu(A,\theta)+\delta_\Lambda\hat{A}_\mu(A,\theta)
=\hat{A}_\mu (A+\delta_\lambda A, \theta).
\end{eqnarray}
On the left, the transformation is performed on $\hat{A}_\mu$. The result should be the noncommutative gauge potential that is associated with $A$ after an usual U(1) transformation, i.e., $A+\delta_\lambda A$ on the right. From Eq.~\eqref{eq:A-transform}, one obtains
\begin{align} \label{eq:A-equivalence}
\nonumber
&\hat{A}_\mu (A+\delta_\lambda A, \theta)
-\hat{A}_\mu(A,\theta)
=\partial_\mu\Lambda 
+i[\Lambda, \hat{A}_\mu]_\star
\\
\Rightarrow~
&\delta_\lambda\hat{A}_\mu (A, \theta)
=\partial_\mu\Lambda 
+i[\Lambda, \hat{A}_\mu]_\star.
\end{align}
Similarly, the noncommutative and commutative matter fields are related by
\begin{align} \label{eq:field-equivalence}
\nonumber
&\Psi(\psi, \theta)+\delta_\Lambda \Psi(\psi,\theta)
=\Psi (\psi+\delta_\lambda \psi, \theta)
\\ \nonumber
\Rightarrow~
&\Psi (\psi+\delta_\lambda \psi, \theta)
-\Psi(\psi, \theta)
=\delta_\Lambda \Psi(\psi,\theta)
\\
\Rightarrow~
&\delta_\lambda \Psi (\psi, \theta)
=i\Lambda\star \Psi.
\end{align}

In order to map or approximate a noncommutative field theory by a commutative field theory, Eqs.~\eqref{eq:Lambda-eq}, \eqref{eq:A-equivalence}, and~\eqref{eq:field-equivalence} need to be solved consistently. This is achieved by assuming the following functional forms for the noncommutative fields and gauge parameters~\cite{Seiberg-Witten}:
\begin{align}
\Lambda
&=\Lambda(\lambda, A, \theta),
\\
\hat{A}
&=\hat{A}(A, \theta),
\\
\Psi
&=\hat{\Psi}(\psi, A, \lambda, \theta).
\end{align}
Then, the noncommutative fields are expanded order by order in $\theta$ such that one recovers the commutative fields as $\theta \rightarrow 0$:
\begin{align}
\Lambda
&=\Lambda_\lambda^{(0)}+\Lambda^{(1)}_\lambda+\Lambda^{(2)}_\lambda+\cdots,
\\
\hat{A}_\mu
&=A_\mu+A_\mu^{(1)}+A_\mu^{(2)}+\cdots,
\\
\Psi
&=\psi+\psi^{(1)}+\psi^{(2)}+\cdots.
\end{align} 
The symbol $\Lambda^{(k)}_\lambda$ denotes a term in the $k$-th order of $\theta$. For $k=0$, we have $\Lambda_\lambda^{(0)}=\lambda$. For $k>0$, $\Lambda^{(k)}_\lambda$ is a function of $\lambda$, $A_\mu$, and their temporal and spatial derivatives. In principle, Eqs.~\eqref{eq:Lambda-eq}, \eqref{eq:A-equivalence}, and~\eqref{eq:field-equivalence} can then be solved order by order in $\theta$. Nevertheless, the task turns out to be very challenging when one considers higher order terms. In Appendix \ref{app:2nd-SW}, we discuss some technical and mathematical results with respect to the SW map, which are important to our work. 

\subsubsection{First order solution in $\theta$}

To illustrate the systematic procedures in determining the SW map, we discuss in detail the first-order SW map here. It starts by deducing $\Lambda_\lambda^{(1)}$ since Eq.~\eqref{eq:Lambda-eq} only involves $\Lambda$ and $\lambda$. By expanding Eq.~\eqref{eq:Lambda-eq} up to the first order in $\theta$, we obtain
\begin{eqnarray}
\delta_\alpha \Lambda_\beta^{(1)}
-\delta_\beta \Lambda_\alpha^{(1)}
=i[\alpha, \beta]_\star
=-\theta^{\mu\nu}\partial_\mu\alpha\partial_\nu\beta.
\end{eqnarray}
An inhomogeneous solution is given by~\cite{BRST-1, Zumino2002, Picariello2002, Fidanza, JMSSW2001}
\begin{eqnarray} \label{eq:Lambda-1}
\Lambda_\lambda^{(1)}
=-\frac{1}{2}\theta^{\mu\nu}A_\mu\partial_\nu\lambda.
\end{eqnarray}
Note that the SW map is not unique. One can always add a homogeneous solution to $\Lambda_\lambda^{(1)}$~\cite{Zumino2002, BRST-1}.

For the gauge potential, performing a similar series expansion for Eq.~\eqref{eq:gauge-equivalence} and retaining terms up to the first order in $\theta$ gives
\begin{align} \label{eq:1eq-A}
\nonumber
&\delta_\lambda\left[A_\mu+A_\mu^{(1)}\right]
=\partial_\mu\left[\lambda+\Lambda^{(1)}_\lambda\right]
-i\left(i\theta^{\alpha\beta}\partial_\alpha A_\mu \partial_\beta\lambda\right)
\\
\Rightarrow
~&\delta_\lambda A_\mu^{(1)}
=\partial_\mu \left(-\frac{1}{2}\theta^{\alpha\beta}A_\alpha\partial_\beta\lambda\right)
-\theta^{\alpha\beta}\partial_\beta A_\mu \partial_\alpha\lambda.
\end{align}
A possible inhomogeneous solution is~\cite{BRST-1, Zumino2002, Picariello2002, Fidanza, JMSSW2001}
\begin{eqnarray} \label{eq:A1}
A_\mu^{(1)}
=-\frac{1}{2}\theta^{\alpha\beta}
A_\alpha\left(\partial_\beta A_\mu + F_{\beta\mu}\right).
\end{eqnarray}
Here, $F_{\beta\mu}=\partial_\beta A_\mu - \partial_\mu A_\beta$ is the field tensor. The above result can be rewritten as the sum of total derivatives (i.e., boundary terms in the action) and a term in the Chern-Simons form~\cite{Picariello2002}:
\begin{align}
\nonumber
A_\mu^{(1)}
=&-\frac{1}{2}\partial_\beta\left(\theta^{\alpha\beta}A_\alpha A_\mu\right)
\\
&+\frac{1}{4}\theta^{\alpha\beta}
\left(
A_\mu F_{\beta\alpha}
+A_\beta F_{\alpha\mu}
+A_\alpha F_{\mu\beta}
\right).
\end{align}
In 3D, the last term for $A_0^{(1)}$ is actually the CS term. The term vanishes when $\mu\neq 0$ in 3D but does not vanish in higher dimensions:
\begin{eqnarray} \label{eq:A1_D+CS}
A_{\mu,\text{3D}}^{(1)}
=-\frac{1}{2}\partial_\beta\left(\theta^{\alpha\beta}A_\alpha A_\mu\right)
-\frac{\hbar}{2}\delta_{\mu, 0}\epsilon^{\alpha\beta\nu}A_\alpha \partial_\beta A_\nu.
\end{eqnarray}

Note that in general, setting $\hat{A}_\mu = 0$ does not exclude the possibility of the corresponding commutative gauge field $A_\mu$ being nonzero under the SW map. When we go to the phase space, this allows us to set the noncommutative fields $\hat{A}_{p_i} = \hat{A}_{x_i} = 0$ for simplicity.

Lastly, we consider the first-order expansion for the matter field using Eq.~\eqref{eq:field-equivalence}:
\begin{align}
\nonumber
&\delta_\lambda\left[\psi+\psi^{(1)}\right]
=i\left[\lambda+\lambda^{(1)}\right]
\left(1+\frac{i}{2}\theta^{\mu\nu}\overleftarrow{\partial_\mu}
\overrightarrow{\partial_\nu}\right)
\left[\psi+\psi^{(1)}\right]
\\
\Rightarrow~
&\delta_\lambda \psi^{(1)}
=i\lambda\psi^{(1)}+i\lambda^{(1)}\psi
-\frac{1}{2}\theta^{\mu\nu}\partial_\mu\lambda\partial_\nu\psi.
\end{align}
Since $\psi^{(1)}$ appears on both sides, $\psi^{(1)}$ must be a product between an expression and $\psi$. After the action of $\delta_\lambda$, the expression should cancel the remaining two terms on the right. This suggests a possible inhomogeneous solution~\cite{JMSSW2001, Ulker2008},
\begin{eqnarray}
\psi^{(1)}=-\frac{1}{4}\theta^{\mu\nu}A_\mu\left(2\partial_\nu-iA_\nu\right)\psi.
\end{eqnarray}

\section{Action for phase-space fermion liquid in arbitrary dimensions}
\label{sec:action}

In this section, we give the general form for the action of an incompressible fermion liquid in phase space in $d$ spatial dimensions. In subsequent sections we will develop the theory for the physically relevant cases of one, two, and three dimensions.

Assuming the usual noncommutativity relations (Eq.~\eqref{eq:NCphaseSpace}) for the position $X_i$ and momentum $P_i$ variables, we propose the following Euclidean action for the incompressible liquid in phase space,
\begin{equation}
\label{eq:generalAction}
    S = \int d^{2d+1} x\ \Bar{\Psi}\star\left(\frac{\partial \Psi }{\partial \tau} - i\hat{A}_0\star \Psi\right) + \cdots
\end{equation}
where the ellipses include terms of quartic and higher order in the fermions. In writing the above action, we have already traded the noncommutativity in the coordinates for the Moyal-Weyl star product. The (classical) phase space measure is
\begin{equation}
    d^{2d+1}x = d\tau\prod_{i = 1}^d dx_i dp_i.
\end{equation}
Incompressibility of the fermion liquid implies that the density operator $\Bar{\Psi}\star\Psi$ and its commutative counterpart (zeroth order in the Seiberg-Witten map) $\rho \equiv \bar{\psi}\psi$ have {\em fixed} expectation values and their fluctuations (and other excitations) gapped out in the bulk, and the low-energy excitations only live on the boundaries. Thus, we write the density in phase-space as a sum of the mean density and fluctuations,
\begin{equation}
\label{eq:incompressibility}
    \bar{\psi}\psi \approx \langle \bar{\psi}\psi\rangle + \delta \rho = \bar{\rho} + \delta\rho
\end{equation}
where $\bar{\rho}$ is given by 
\begin{equation}
    \bar{\rho} = \frac{1}{(2\pi\hbar)^d},
\end{equation}
and fluctuations above it can be integrated out perturbatively.

In a fully translation-invariant system the fermion liquid has a boundary in the momentum directions corresponding to the Fermi surface, and no boundary in real space because periodic boundary conditions need to be imposed in real-space directions. In reality, on the other hand, there \textit{is} a real-space boundary which breaks translation invariance. The advantage of our phase-space formalism is that boundaries in real space and momentum space, which are both parts of phase-space boundaries, can be treated on the same footing. In order to contain the liquid, there must be a confining potential in all directions in $\hat{A}_0(x_i, p_i, \tau)$, which is nothing but the noninteracting Hamiltonian for the fermion liquid:
\begin{equation}
    \hat{A}_0(\textbf{x}, \textbf{p}, \tau) = \hat{\epsilon}(\textbf{p}) - \mu + \hat{V}_0(\textbf{x}, \textbf{p}),
\end{equation}
where $\epsilon(\textbf{p})$ is the band dispersion and $\hat{V}$ is the external potential (which could in principle be momentum dependent). So, in our phase-space formulation the kinetic (band) energy is nothing but a confining potential in momentum space, which is treated on equal footing with real-space potential.
For simplicity, we set $\hat{A}_p=\hat{A}_x=0$.

Since a star product between two functions under the integral can be replaced by the ordinary product if boundary terms can be neglected~\cite{RMP-noncommutative, Szabo, Wohlgenannt, Barbon}, the action can be simplified to
\begin{equation}
\label{eq:action-ArbD}
    S = \int d^{2d + 1}x \left[
\bar{\Psi}\left(\frac{\partial \Psi}{\partial\tau}
-i\hat{A}_0\star \Psi
\right)\right] + \ldots
\end{equation}
The above form will be the starting point for our discussions from hereon.

\section{Emergence of the Chern-Simons term in 1D}
\label{sec:1D-FL}

Let us study one-dimensional fermion liquids using the phase-space formalism. We know that the topological response is captured by the CS action or ETS anomaly in Eq.~\eqref{eq:1D}. As we will demonstrate below, this term emerges naturally by employing the SW map to the noncommutative field theory formulated for the incompressible liquid in the (2+1)-dimensional (phase) space time. This trick was first employed in the recent studies of composite fermion theory restricted to the lowest Landau level~\cite{Dong2020}. The main difference is that the CS term we will obtain is for the probe field $A_\mu$, but not the emergent gauge field. Thus, it indeed governs the anomaly and topological responses of the system under the external electromagnetic field. Using the zeroth order term from the SW map in Eq.~\eqref{eq:action-ArbD} , we get
\begin{align}
\label{eq:S-1:1d}
S
=&\int d^3\bm{x}
\left[
\bar{\psi}\left(\partial_\tau-iA_0\right)\psi\right] + \cdots
\end{align}

Hereafter, we assume $\Psi$ and $\psi$ are time independent. As $\bar{\psi}\psi \approx \bar{\rho}\sim \hbar^{-1}$, the term written explicitly in Eq.~\eqref{eq:S-1:1d} is of the order $\mathcal{O}(1/\hbar)$.

Next, we consider the first-order $\theta$ expansion for $\hat{A}_\mu$ in Eq.~\eqref{eq:A1}. The zeroth component is given by
\begin{eqnarray}
\hat{A}_0^{(1)}
=-\frac{1}{2}\theta^{\mu\nu}A_\mu\left(\partial_\nu A_0+F_{\nu 0}\right).
\end{eqnarray}

Plugging in the incompressibility assumption Eq.~\eqref{eq:incompressibility}, we have
\begin{align}
\nonumber
S
\equiv&\ S_{-1} + S_{0} + \ldots
\\
=&\int d^3\bm{x}
\left[-iA_0\right]\left(\bar{\rho}+\delta\rho\right) \nonumber
\\
&+\int d^3\bm{x}~
\frac{i}{2}\theta^{\mu\nu}A_\mu\left(\partial_\nu A_0+F_{\nu 0}\right)
\left(\bar{\rho}+\delta\rho\right) + \ldots
\end{align}
Using integration by parts or Eq.~\eqref{eq:A1_D+CS} directly, one obtains
\begin{eqnarray}
S_0
=\frac{i\hbar{\bar{\rho}}}{2}
\int d^3\bm{x}~
\epsilon^{\mu\nu\sigma}A_\mu\partial_\nu A_\sigma
+S_{0,B} + \ldots
\end{eqnarray}
where $S_{0,B}$ denotes the boundary term in Eq.~\eqref{eq:A1_D+CS}. 

Finally, using $\bar{\rho} = 1/(2\pi\hbar)$ and ignoring the terms proportional to $\delta \rho$, we get
\begin{align}
S_0 - S_{0, B}
=\frac{i}{2(2\pi)}
\int d^3\bm{x}~
\epsilon^{\mu\nu\sigma}A_\mu\partial_\nu A_\sigma.
\end{align}
We thus obtain the desired Chern-Simons term, and it is in this sense we have a quantum Hall liquid in phase space. For the remaining terms, we have 
\begin{eqnarray} \label{const-term}
S_{-1}
=\int d^3\bm{x}
\left[-iA_0\right] \delta\rho
\end{eqnarray}
and the boundary term from integration by parts is
\begin{eqnarray}
S_{0,B}
=\frac{i\hbar{\bar{\rho}}}{2}
\int d^3\bm{x}~
\left[\partial_2\left(A_0 A_1\right)-\partial_1\left(A_0 A_2 \right)\right].
\end{eqnarray}

The existence of boundary terms begets the question of gauge invariance of the effective actions. Of course, the starting action Eq.~\eqref{eq:action-ArbD} is explicitly gauge invariant under a U(1) gauge transformation in phase space. The full SW map preserves this gauge invariance. The constraint $\partial_\tau \Psi = 0$ is a gauge choice, which explains the apparent lack of gauge invariance at order $\mathcal{O}(1/\hbar)$ in $S_{-1}$. Additionally, $S_{0,B}$ is exactly canceled by a term that one gets upon relaxing incompressibility,
\begin{align}
\begin{split}
    S_{0, B2} &= \int d^3 x\  [-iA_0](\bar{\psi}\psi^{(1)} + \bar{\psi}^{(1)}\psi) \\
    &= \frac{i}{2}\int d^3 x\ \theta^{\mu\nu}\left[A_0A_\mu \partial_\nu(\bar{\psi}\psi)\right]\\
    &= - \frac{i\hbar}{2}\int d^3 x\  \bar{\psi}\psi\left[\partial_2(A_0A_1) - \partial_1(A_0A_2)\right]
\end{split}
\end{align}
which becomes precisely $-S_{0,B}$ upon assuming $\bar{\psi}\psi \approx \bar{\rho}$. Finally, we are left with the CS term, whose gauge invariance must be imposed by additional boundary degrees of freedom corresponding to the edge excitations of the quantum Hall liquid, as expected.

\section{2D fermion liquid as 4D phase-space quantum Hall liquid}
\label{sec:2D-FL}

Next, we specialize to two spatial dimensions, where the action, Eq.~\eqref{eq:action-ArbD}, is integrated over $(2+2+1)$-D space with
$d^5\bm{x}=d\tau dx dp_x dy dp_y $. Again, the Fermi surface $(p_x^F(\theta), p_y^F(\theta))$ defines the boundary in the momentum directions. 

Following the previous example on one-dimensional space, we assume $\Psi$ and $\psi$ are time-independent. The mean density is now $\bar{\rho} = 1/(2\pi\hbar)^2$. The leading term in the effective action for the fermion liquid is thus of order $\mathcal{O}(1/\hbar^2)$: 
\begin{eqnarray}
\label{eq:secondOrderzeroth}
S_{-2}
=\frac{1}{(2\pi\hbar)^2}\int d^5\bm{x}
\left[-iA_0\right]
+\cdots
\end{eqnarray}

\subsection{Order-by-order expansion in $\theta$}
\subsubsection{Contribution from first order expansion for $\hat{A}_0$}
The first subleading term scales as $\theta^{-1}$,
\begin{align}
\begin{split}
S_{-1}
&=\int d^5\bm{x} \left[-iA_0^{(1)}\bar{\psi}\psi\right]\\
&=\frac{i\bar{\rho}}{2}\theta^{\alpha\beta} \int d^5\bm{x} 
\left[A_\alpha\left(\partial_\beta A_0 + F_{\beta 0}\right)\right]
\end{split}
\end{align}
upto contributions from the fluctuations $\delta \rho$. We define the coordinate system for the phase space with imaginary time as 
$(\tau, x, p_x, y, p_y)$. Here, $\tau$ is the zeroth coordinate. Thus, the antisymmetric tensor 
$\theta$ takes a symplectic form,
\begin{eqnarray}
\theta^{12}=\theta^{34}=1
,~
\theta^{21}=\theta^{43}=-1
,~
\text{and zero otherwise}
\end{eqnarray}
Using the previous trick of integration by parts [or Eq.~\eqref{eq:A1_D+CS}] twice, we have
\begin{align} \label{eq:S-15D}
\begin{split}
S_{-1}
&=\frac{i\bar{\rho}}{2}\theta^{\alpha\beta} \int d^5\bm{x} 
\left[A_\alpha\left(\partial_\beta A_0 + F_{\beta 0}\right)\right]
\\
&=\frac{i\hbar{\bar{\rho}}}{2}
\left[
\left.\int d^5\bm{x}~
\epsilon^{\mu\nu\sigma}A_\mu\partial_\nu A_\sigma
\right|_{\substack{~~(\mu,\nu,\sigma)\\=(0,1,2)}}\right.
\\
&+\left.\left.\int d^5\bm{x}~
\epsilon^{\mu\nu\sigma}A_\mu\partial_\nu A_\sigma
\right|_{\substack{~~(\mu,\nu,\sigma)\\=(0,3,4)}}
\right]
+S_{-1,B}
\end{split}
\end{align}
This is similar to having two decoupled Landau levels in $(x,p_x)$ and $(y,p_y)$~\cite{Yang-PRA2020}. Note that $S_{-1}$ is not a genuine Chern-Simons term in the five-dimensional spacetime, but the two three-dimensional subspaces $(\tau, x, p_x)$ and $(\tau, y, p_y)$. $S_{-1,B}$ collects the boundary terms from the two lower-dimensional CS forms.

The CS 3-forms in $S_{-1}$ are ``nontopological" in the sense that their coefficients are not quantized in 4+1-D phase space. However, this does not mean that their forms are arbitrary. In fact, $S_{-1}$ is invariant under any canonical transformations in phase space. Under a canonical transformation $$(x, y, p_x, p_y) \rightarrow (x', y', p'_x, p'_y),$$
the phase space measure remains invariant due to the symplectic condition. For the gauge fields, $A_0$ remains invariant, but the remaining $A_\mu$ transform as covariant vectors
\begin{equation}
    A'_\mu = \sum_{\nu \ne 0} \frac{\partial x'^\nu }{\partial x^\mu}A_\nu.
\end{equation}
Let us consider the set of terms 
\begin{equation}
    A_0(\partial_1 A_2 - \partial_2 A_1 + \partial_3 A_4 - \partial_4 A_3). 
\end{equation}
After the canonical transformations, they remain invariant due to the invariance of the Poisson bracket $\{x'^\mu, x'^\nu\}$ between canonical coordinates:
\begin{multline}
\label{eq:CanonicalInvariance}
    A'_0\left(\sum_{\nu, \sigma \neq 0} \frac{\partial x'^\nu}{\partial x^1}\frac{\partial x'^{\sigma}}{\partial x^2} - \frac{\partial x'^\nu}{\partial x^2}\frac{\partial x'^{\sigma}}{\partial x^1} \right)\frac{\partial A'_\sigma}{\partial x'_\nu} \\
   + A'_0\left(\sum_{\nu, \sigma \neq 0} \frac{\partial x'^\nu}{\partial x^3}\frac{\partial x'^{\sigma}}{\partial x^4} - \frac{\partial x'^\nu}{\partial x^4}\frac{\partial x'^{\sigma}}{\partial x^3} \right)\frac{\partial A'_\sigma}{\partial x'_\nu} \\
   = A'_0 \sum_{\nu, \sigma \neq 0}\{x'^{\nu}, x'^{\sigma}\}\frac{\partial A'_\sigma}{\partial x'_\nu} \\= A'_0(\partial'_1 A'_2 - \partial'_2A'_1 + \partial'_3A'_4 - \partial'_4 A'_3).
\end{multline}

The same holds true for the remaining terms in the CS 3-form as well. Thus, the total contribution to the physical responses from the nontopological terms are invariant under arbitrary canonical transformations in phase space. 

\subsubsection{Contribution from second order expansion for $\hat{A}_0$}

Finally, let us consider the contribution from $A_0^{(2)}$,
\begin{eqnarray}
S_0 = \int d^5\bm{x} \left[-iA_0^{(2)}\bar{\psi}\psi\right].
\end{eqnarray}
In Appendix~\ref{app:2nd-SW}, we show that in 5D phase space,  $A_0^{(2)}$ can be rewritten as the sum of a Chern-Simons form and boundary terms (Eq.~\eqref{eq:5D-A2}). A direct substitution of the result gives
\begin{equation}
\label{eq:SecondOrderwMatter}
S_{0} - S_{0, B}
=\frac{-i\bar{\rho}\hbar^2}{6}\int d^5\bm{x}~ 
\epsilon^{\alpha\beta\sigma\rho\nu}
A_\alpha \partial_\beta A_\sigma \partial_\rho A_\nu \\
\end{equation}

where $S_{0,B}$ is a surface term contributed by the first line in Eq.~\eqref{eq:5D-A2}. Thus, we arrive at one of our main results, the first term in Eq.~\eqref{eq:SecondOrderwMatter} is the correct dimensionless CS 5-form with the correct prefactor $1/[6(2\pi)^2]$.

The topological CS term describes several contributions to responses in the 2D EFL \cite{Qi2015, ETS}, especially the unquantized part of the anomalous Hall response \cite{AHEReview} in the presence of nonzero Berry curvature $F_{p_x, p_y}$\footnote{The momentum-space gauge fields $A_{p_x}, A_{p_y}$ correspond to the Berry connection \cite{Qi2015}.}. Additionally, if we view the current as being generated by Bloch wavepackets $\phi_{\textbf{R}, \textbf{K}}(x,y)$, we also obtain the response due to mixed space Berry curvatures~\cite{DiXiao2010, Sundaram1999, Randeria2024}, such as
\begin{equation}
 F_{x_i,p_j} = \partial_{x_i} A_{p_j} - \partial_{p_j}A_{x_i}   
\end{equation} 
which can be activated for electronic wavepackets moving in the presence of a topologically non-trivial magnetic structure.

\subsection{Semiclassical terms in the response}
 
An upshot of our treatment of the phase-space quantum Hall liquid is that in addition to the topological CS term in 4+1-D phase space proposed by ETS, we also get the lower-dimensional (and nontopological) CS forms [Eq.~\eqref{eq:S-15D}]. The CS 3-forms ($\mathrm{CS}_3$) generate the group velocity term in the current,
\begin{equation}
\label{eq:JxCS3}
J_{x_i}^{\mathrm{CS}_3} \propto \int dp_x dp_y\  \delta \rho\  \partial_{p_i}A_0.
\end{equation}
As noted earlier, the $p$-dependent part of $A_0$ is the dispersion $\epsilon(p)$ and thus $\partial_{p_i}A_0 = \partial_{p_i}\epsilon$ is the group velocity.  Adopting the electronic wave-packet point of view, we also get its momentum-space equivalent
\begin{equation}
\label{eq:JpCS3}
    J_{p_i}^{\mathrm{CS}_3} \propto \frac{dK_i}{dt} = -\partial_{x_i}A_0,
\end{equation}
which is nothing but Newton's second law. These semiclassical terms were absent from earlier works focusing only on the topological term. In fact, the semiclassical response will always be generated for the ersatz Fermi liquid in any space-time dimension. Consider the EFL in $d$ spatial dimensions. The first-order SW map for $\hat{A}_0$ will always yield a set of terms like Eq.~\eqref{eq:S-15D}, for each set of conjugate pairs $(x_i, p_{x_i})$. The semiclassical contributions are essentially the Hamilton-Jacobi equations for a free electron moving in the presence of an electric field, and thus are invariant under arbitrary canonical transformations, as we already saw in Eq.~\eqref{eq:CanonicalInvariance}. This is what fixes the form of $S_{-1}$.

\section{Fermion liquids in 3D and beyond}
\label{sec:generalization}

The main obstacle to generalizing our results to arbitrary $d$ is that it is difficult to verify if one always gets the topological CS term in the Seiberg-Witten map, though this has been conjectured in earlier literature, such as Ref. \cite{Picariello2002}. There is a recursive solution to the SW map at all orders (Ref. \cite{Fidanza} and Appendix~\ref{app:2nd-SW}), which ensures that the topological term Eq.~\eqref{eq:general D} can only arise at order $d$ on the SW map, but not earlier. This implies one needs to go to higher orders in the SW map in higher dimensions in order to obtain the topological CS term, which quickly becomes quite challenging. For the case of three spatial dimensions, we were able to verify the conjecture (Appendix \ref{sec:3rdOrder}) using an approach analogous to the proof in two dimensions. The main result of Appendix \ref{sec:3rdOrder} is that the third-order term $A_0^{(3)}$ in the SW map for the gauge field can be rewritten as a CS term in 6+1-D phase space, upto total derivatives. Thus, for the physically relevant cases of $d = 1,2,3$ the topological CS term is in fact derived from the SW map.

\begin{figure} [htb]
\centering
\includegraphics[width=3.5in]{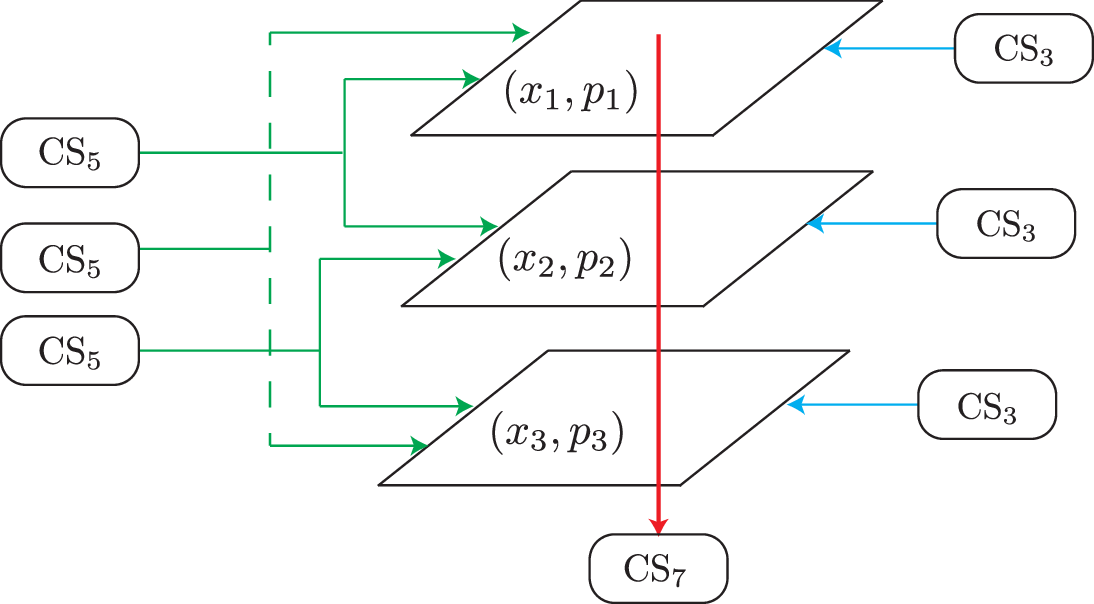}
\caption{Generation of different CS actions in the six-dimensional phase space with imaginary time, and the corresponding subspaces with smaller odd dimensions under the SW map. Note that there is only a single imaginary time, $\tau$.}
\label{fig:SW-map-generalization}
\end{figure}

For the three-dimensional fermion liquids, similar to the previous cases, the lowest order expansion consists of several terms (Fig. \ref{fig:SW-map-generalization}),
\begin{eqnarray}
S=S_{-3} + S_{-2}+S_{-1}+S_0+\cdots,
\end{eqnarray}
where
\begin{align}
S_{n-3}
&=\int d^7\bm{x}\left[-iA_0^{(n)}\right]\bar{\psi}\psi
\end{align}
for $n \in (0,1,2,\ldots)$. By generalizing Eq.~\eqref{eq:S-15D} and assuming incompressibility, we see that $S_{-2}$ consists of three terms that describe the CS 3-form in each set of conjugate variables $(x,p_x)$, $(y,p_y)$, and $(z,p_z)$. Similarly, $S_{-1}$ consists of three terms that describe the CS 5-form in separate pairs formed by any two sets of conjugate variables $(x,p_x)$, $(y,p_y)$, and $(z,p_z)$, with a boundary term:
\begin{multline}
    S_{-1}[A] =  \frac{-i\Omega_{z,p_z}}{6(2\pi)^3\hbar}\left.\int d^5x\ \epsilon^{\mu\nu\sigma\rho\delta}A_\mu \partial_\nu A_\sigma \partial_\rho A_\delta\right|_{\substack{~~(\mu,\nu,\sigma, \rho, \delta)\\=(0,1,2,3,4)}}\\
    + (x\rightarrow y\rightarrow z) + (x\rightarrow z\rightarrow y) + S_{-1,B} + \mathcal{O(\delta\rho)}
\end{multline}
where $\Omega_{x_i,p_{x_i}}$ is the volume in phase space along the specified directions and $S_{-1,B}$ collects all the boundary terms.

At order $S_0$ (contribution of the third order SW term $A_0^{(3)}$), we have the topological CS term
\begin{equation}
\label{eq:CS7}
    S_0 - S_{0, B}= \frac{i}{(2\pi)^34!}\int A\wedge dA \wedge dA\wedge dA
\end{equation}
with the correct quantized coefficient for 7D (phase) space-time.

Again, like the 2D scenario, we obtain the nontopological terms $S_{-1}$ and $S_{-2}$ which describe physically relevant responses of the fermion liquids. $S_{-2}$ contains three sets of CS 3-forms for each pair of conjugate variables $(x_i, p_i)$ and generates the semiclassical responses Eqs.~\eqref{eq:JxCS3} and~\eqref{eq:JpCS3}. The set of CS 5-forms $S_{-1}$ is responsible for generating the anomalous Hall responses in all three spatial currents $J_x, J_y,$ and $J_z$ when the Fermi surface encloses nonzero Berry flux
\begin{equation}
    \sigma^{CS_5}_{x_i, x_j} \propto \int dp_i dp_j\ F_{p_i, p_j}.
\end{equation}
The nonquantized part of the anomalous Hall response in the equilibrium current is yet another important contribution that cannot be obtained from the topological CS term [Eq.~\ref{eq:CS7}] that yields the ETS anomaly. Thus, our approach unifies all possible responses of the fermion liquid to real- or momentum-space gauge fields.

\section{Concluding Remarks} 
\label{sec:conclusion}

We have shown that fermion liquids with Fermi surfaces satisfying the Luttinger sum rule can be viewed as incompressible liquids in phase space, which are topologically non-trivial in the way similar to quantum Hall liquids. Taking into consideration the noncommutativity of phase space, we employed the Seiberg-Witten map to derive the effective action governing their responses to external probing gauge fields. In addition to the topological term that captures the ETS anomaly, the phase-space formalism also yields lower-order terms that reproduce semiclassical and other nonquantized responses in 2D and 3D. From the viewpoint of noncommutative field theory, our work also presents an important technical advancement. It has been a longstanding conjecture \cite{Picariello2002} that CS forms arise at all orders in the SW map. We prove that this is indeed true, up to order 3. In Appendix \ref{app:higherOrders}, we sketch a proof by induction for the higher-order terms.

Our work unifies the description of Fermi and non-Fermi liquids, as they share the {\em same} bulk (in phase space) topological properties. Their difference, on the other hand, lies at the phase space boundary, including the Fermi surface. This is similar to what happens in quantum Hall liquids with a specific bulk topological order, where edge physics can still be different due to non-universal physics like edge reconstruction. As a concrete example, the edge of a $\nu=1$ quantum Hall liquid is a Fermi liquid without reconstruction, but becomes a Luttinger liquid with reconstruction. Our work further unifies the description of Fermi surface (or momentum-space boundary) with real-space boundary, as they are both parts of phase-space boundary. Strictly speaking, Fermi surfaces are {\em not} sharp in the presence of disorder and/or a real-space boundary. As both break translation symmetry, one can only talk about phase-space boundary in such cases. This suggests the possibility of probing Fermi surface physics at the real-space boundary of the system, in a way similar to edge physics of quantum Hall liquids.

In a way our work also unifies the description of metals and (band or Mott) insulators, which are normally considered the exact opposite of each other. The similarity between them, in the context of the present paper, lies in their incompressibility, one in phase space and the other in real space. In fact, insulators are also incompressible in phase space, except their boundaries are confined to real space (in other words, they have no Fermi surface!). From this viewpoint the difference between metal and insulator lies in the (phase space) boundary, with insulators being simpler because their phase-space boundaries are orthogonal to the momentum space directions.

This line of work will likely lead to new insights into high-dimensional bosonization (of the Fermi surface), the formulation of which has not been entirely satisfactory. A promising formulation has appeared recently in Ref. ~\cite{Son-PRR2022} and subsequent works \cite{Huang_2024, Ye-2024, Ye-2024-2}. Our perspective, namely, treating the Fermi surface as the boundary of an incompressible liquid in phase space, is complementary to this approach.

It is tempting to generalize our approach to study interacting quantum Hall liquid in phase spaces that are topologically distinct from those studied here, and explore its physical consequences. In particular, one can consider a fractional version of the incompressible liquids with higher levels of CS term in their effective action. Such liquids satisfy a fractional version of the Luttinger sum rule, in a way similar to fractional quantum Hall liquids with a fractional bulk filling factor. Due to the quantization of the level of the CS term, the phase space density of such liquids must be a rational fraction of that dictated by the Luttinger sum rule, the simplest version of which is $1/m$, with $m$ being an integer. This corresponds to the Laughlin sequence in fractional quantum Hall effect \cite{Agarwal-Karabali-2025}. It is not immediately clear, however, whether $m$ has to be odd for $d \ge 2$. We leave these issues to be addressed in future works.
\begin{acknowledgments}

The authors acknowledge helpful conversations with T. Senthil. J.P. acknowledges the support from Florida State University through the FSU Quantum Postdoctoral Fellowship. This research is supported by the National Science Foundation Grant No. DMR-2315954. The portion of the work conducted at Northeastern University was partially supported by NSF CAREER Award No. DMR-2439118. Most of this work was performed at the National High Magnetic Field Laboratory, which is supported by National Science Foundation Cooperative Agreement No. DMR-2128556, and the State of Florida.

\end{acknowledgments}

\appendix

\onecolumngrid

\section{Technical details of the second-order Seiberg-Witten map}
\label{app:2nd-SW}

In the main text, we have demonstrated explicitly how to obtain the first-order SW map. The result can be used to deduce the second-order SW map, which was used in Sec.~\ref{sec:2D-FL} to derive the Chern-Simons action and different responses of fermion liquids in two dimensions. However, the derivation of second and higher order SW map is tedious and highly technical. In this Appendix, we provide the derivation based on the useful results reported in previous literature~\cite{Picariello2002, Fidanza}. \\

By expanding Eq.~\eqref{eq:Lambda-eq} to second order of $\theta$, we obtain
\begin{eqnarray}
\delta_\alpha\left[\beta+\Lambda_\beta^{(1)}+\Lambda_\beta^{(2)}\right]
-\delta_\beta\left[\alpha+\Lambda_\alpha^{(1)}+\Lambda_\alpha^{(2)}\right]
=-\theta^{\mu\nu}\left[\partial_\mu\alpha +\partial_\mu\Lambda_\alpha^{(1)}\right]
\left[\partial_\nu\beta +\partial_\nu\Lambda_\beta^{(1)}\right]
\end{eqnarray}
Comparing different terms order by order in $\theta$, we have
\begin{align}
&\delta_\alpha\beta-\delta_\beta\alpha
=0,
\\
&\delta_\alpha\Lambda_\beta^{(1)}-\delta_\beta\Lambda_\alpha^{(1)}
=-\theta^{\mu\nu}\partial_\mu\alpha\partial_\nu\beta
\end{align}
The solution to the first-order equation is given in Eq.~\eqref{eq:Lambda-1}. Our main focus here is the second-order equation,
\begin{eqnarray}
\delta_\alpha\Lambda_\beta^{(2)}-\delta_\beta\Lambda_\alpha^{(2)}
=\frac{1}{2}\theta^{\mu\nu}\theta^{\sigma\rho}
\left[
\left(\partial_\mu\alpha\right)\partial_\nu(A_\sigma\partial_\rho\beta)
+\left(\partial_\nu\beta\right) \partial_\mu(A_\sigma\partial_\rho\alpha)
\right]
\end{eqnarray}
A possible inhomogeneous solution is given by~\cite{Picariello2002, Fidanza}:
\begin{eqnarray} \label{eq:Lambda2}
\Lambda_\lambda^{(2)}
=\frac{1}{6}\theta^{\mu\nu}\theta^{\sigma\rho}
A_\nu\left[\partial_\mu\left(A_\rho\partial_\sigma\lambda\right)
-F_{\rho\mu}\partial_\sigma\lambda\right]
\end{eqnarray}
Furthermore, Ref.~\cite{Picariello2002} pointed out explicitly that $\Lambda_\lambda^{(2)}$ can be written as
\begin{eqnarray}
\Lambda_\lambda^{(2)}
=\frac{1}{6}\partial_\mu
\left(\theta^{\mu\nu}\theta^{\sigma\rho}
A_\nu A_\rho\partial_\sigma\lambda
\right)
-\frac{1}{12}\theta^{\mu\nu}\theta^{\sigma\rho}
\left(
A_\rho F_{\mu\nu}+A_\mu F_{\nu\rho}+A_\nu F_{\rho\mu}
\right)\partial_\sigma\lambda
\end{eqnarray}
In $4+1$-D, the second term can be rewritten in a Chern-Simons form. Hence, 
\begin{eqnarray} \label{eq:L2-CS}
\Lambda_\lambda^{(2)}
=\frac{1}{6}\partial_\mu
\left(\theta^{\mu\nu}\theta^{\sigma\rho}
A_\nu A_\rho\partial_\sigma\lambda
\right)
-\frac{1}{6}\hbar^2\epsilon^{\mu\nu\sigma\rho}A_\rho 
\partial_\mu A_\nu \partial_\sigma\lambda
\end{eqnarray}
The above rewriting is achieved by using the following identity,
\begin{eqnarray} \label{eq:identity1}
\theta^{\mu\nu}\theta^{\sigma\rho}
\left(
A_\rho F_{\mu\nu}+A_\mu F_{\nu\rho}+A_\nu F_{\rho\mu}
\right)\partial_\sigma f
=\hbar^2\epsilon^{\mu\nu\sigma\rho}A_\rho F_{\mu\nu}\partial_\sigma f
=2\hbar^2\epsilon^{\mu\nu\sigma\rho}A_\rho \partial_\mu A_\nu \partial_\sigma f,
\end{eqnarray}
where $f$ can be any function. Another useful identity for later discussion is
\begin{eqnarray} \label{eq:identity2}
\theta^{\mu\nu}\theta^{\sigma\rho}
\left(\partial_\mu A_\nu F_{\sigma\rho}
+\partial_\mu A_\sigma F_{\rho\nu}
+\partial_\mu A_\rho F_{\nu\sigma}
\right)
=2\hbar^2\epsilon^{\mu\nu\sigma\rho}\partial_\mu A_\nu\partial_\sigma A_\rho
\end{eqnarray}
Note that these identities only work in 4+1-D with the Greek letters taking nonzero value. \\

By expanding Eq.~\eqref{eq:gauge-equivalence} to second order in $\theta$, we have
\begin{align}
\delta_\lambda\left[A_\mu+A_\mu^{(1)}+A_\mu^{(2)}\right]
=\partial_\mu\left[\lambda+\Lambda_\lambda^{(1)}+\Lambda_\lambda^{(2)}\right]
-\theta^{\alpha\beta}
\partial_\beta \left[A_\mu+A_\mu^{(1)}\right]
\partial_\alpha \left[\lambda+\Lambda_\lambda^{(1)}\right]
\end{align}
Comparing terms order by order in $\theta$, we have
\begin{align}
&\delta_\lambda A_\mu = \partial_\mu\lambda
\\
&\delta_\lambda A_\mu^{(1)}
=\partial_\mu \Lambda_\lambda^{(1)}
-\theta^{\alpha\beta}\partial_\beta A_\mu \partial_\alpha \lambda
\\
&\delta_\lambda A_\mu^{(2)}
=\partial_\mu \Lambda_\lambda^{(2)}
-\theta^{\alpha\beta}\partial_\beta A_\mu \partial_\alpha \Lambda_\lambda^{(1)}
-\theta^{\alpha\beta}\partial_\beta A_\mu^{(1)} \partial_\alpha \lambda
\end{align}
It is challenging to solve the equation for $A_\mu^{(2)}$. A possible solution is reported in Ref.~\cite{Fidanza}:
\begin{eqnarray} \label{eq:Fidanza-A22}
A_\mu^{(2)}
=\frac{1}{6}\theta^{\alpha\beta}\theta^{\rho\sigma} A_\alpha
\left[
\partial_\beta \left(A_\rho\partial_\sigma A_\mu\right)
+ F_{\beta\rho}\partial_\sigma A_\mu
+2\partial_\beta\left(A_\rho  F_{\sigma\mu}\right)
+2 F_{\beta\rho} F_{\sigma\mu}
\right]
\end{eqnarray}
Notice that other forms for $A_\mu^{(2)}$ have been reported, which differ from each other and Eq.~\eqref{eq:Fidanza-A22} by a total derivative~\cite{JMSSW2001}. It is difficult to determine whether $A_\mu^{(2)}$ (in particular, $A_0^{(2)}$) can be expressed as a sum of boundary terms and a CS 5-form. We will later discuss a recursive solution that allows us to verify such a possibility and generalize to higher dimensions. Using similar procedures, one can derive an expression for $\psi^{(2)}$~\cite{Ulker2008}. The expression is very lengthy, but its exact form is not important in our discussion.

\subsection{Double series expansion, recursive solution, and Chern-Simons term in 5D}
\label{sec:double-expansion}

In the Seiberg-Witten map, one usually expands $\Lambda$, $\hat{A}_\mu$, and $\Psi$ in powers of the noncommutative parameter $\theta^{\alpha\beta}$. Another possible expansion is in a power series of gauge potential. Then, each noncommutative quantity becomes a power series of both $\theta$ and $A$. One writes
\begin{align}
\Lambda
&=\sum_{n,m}\Lambda_\lambda^{(n,m)}\left(A^n, \theta^m\right)
\\
\hat{A}
&=\sum_{n,m} A^{(n,m)}\left(A^n, \theta^m\right)
\end{align}
Here, $A^{(n,m)}$ denotes the term involving a product of $n+1$ $A$ and $\partial A$ and the $m$-th power of $\theta$ and. Using this double expansion, one has the recursive solution,
\begin{align}
\label{eq:recur-lambda}
\Lambda_\lambda^{(n,n)}
&=-\frac{1}{n+1}\theta^{\alpha\beta}A_\alpha^{(n-1,n-1)}\partial_\beta\lambda
\\
\label{eq:recur-A}
A_\mu^{(n,n)}
&=-\frac{1}{n+1}\theta^{\alpha\beta}
A_\alpha^{(n-1, n-1)}\left(\partial_\beta A_\mu+n F_{\beta\mu}\right)
\end{align}
Note that in these recursive solutions, one needs to treat $A_\alpha^{(n-1, n-1)}$ as an operator acting from the left to all terms on their right hand side. This is important in dealing with the differential operators. Now, we will use Eq.~\eqref{eq:recur-A} to rewrite $A_\mu^{(2)}$.

We first use Eq.~\eqref{eq:recur-lambda} and~\eqref{eq:recur-A} to verify with the solutions for 
$\Lambda_\lambda^{(2)}$ and $A_\mu^{(2)}$ in Eqs.~\eqref{eq:Lambda2} and~\eqref{eq:Fidanza-A22}, respectively. In the second order expansion in $\theta$, one only has $\Lambda_\lambda^{(2)}=\Lambda_\lambda^{(2,2)}$ and $A_\mu^{(2)}=A_\mu^{(2,2)}$. The same feature holds for $\Lambda_\lambda^{(1)}$ and $A_\mu^{(1)}$ in the first order expansion in $\theta$~\cite{Fidanza}. Using Eq.~\eqref{eq:recur-lambda},
\begin{align}
\nonumber
\Lambda_\lambda^{(2,2)}
&=-\frac{1}{3}\theta^{\alpha\beta}A_\alpha^{(1,1)}\partial_\beta\lambda
\\ \nonumber
&=-\frac{1}{3}\theta^{\alpha\beta}
\left[-\frac{1}{2}\theta^{\mu\nu}
A_\mu\left(\overrightarrow{\partial_\nu} A_\alpha 
+ F_{\nu\alpha}\right)\right]\partial_\beta\lambda
\\
&=\frac{1}{6}\theta^{\mu\nu}\theta^{\alpha\beta}A_\mu
\left[\partial_\nu \left(A_\alpha \partial_\beta\lambda\right)
 + F_{\nu\alpha}\partial_\beta\lambda\right].
\end{align}
This agrees with Eq.~\eqref{eq:Lambda2} after a redefinition of the indices. Using Eq.~\eqref{eq:recur-A}, we have,
\begin{align}
\nonumber
A_\mu^{(2,2)}
&=-\frac{1}{3}\theta^{\alpha\beta}A_\alpha^{(1,1)}
\left(\partial_\beta A_\mu+2 F_{\beta\mu}\right)
\\ \nonumber
&=-\frac{1}{3}\theta^{\alpha\beta}
\left[-\frac{1}{2}\theta^{\sigma\rho}
A_\sigma\left(\overrightarrow{\partial_\rho} A_\alpha 
+ F_{\rho\alpha}\right)\right]
\left(\partial_\beta A_\mu+2 F_{\beta\mu}\right)
\\
&=\frac{1}{6}\theta^{\sigma\rho}\theta^{\alpha\beta}A_\sigma
\left[\partial_\rho \left(A_\alpha \partial_\beta A_\mu\right)
+ F_{\rho\alpha}\partial_\beta A_\mu 
+2\partial_\rho \left(A_\alpha F_{\beta\mu}\right)
+ 2F_{\rho\alpha}F_{\beta\mu}
 \right].
\end{align}
This is exactly the expression in Eq.~\eqref{eq:Fidanza-A22}. \\ 

Meanwhile, $A_\alpha^{(1,1)}$ has been rewritten as a sum of derivative term and a CS form in Eq.~\eqref{eq:A1_D+CS}. Let us repeat it here with a new set of indices,
\begin{eqnarray}
A_\alpha^{(1,1)}
=-\frac{1}{2}\partial_\rho\left(\theta^{\sigma\rho} A_\sigma A_\alpha\right)
+\frac{1}{4}\theta^{\sigma\rho}
\left(
A_\alpha F_{\rho \sigma}
+A_\rho F_{\sigma \alpha}
+A_\sigma F_{\alpha \rho}
\right)
\end{eqnarray}
Using it and the recursive solution, we obtain:
\begin{align}
\nonumber
A_\mu^{(2,2)}
&=-\frac{1}{3}\theta^{\alpha\beta}
\left[
-\frac{1}{2}\overrightarrow{\partial_\rho}\left(\theta^{\sigma\rho} A_\sigma A_\alpha\right)
+\frac{1}{4}\theta^{\sigma\rho}
\left(
A_\alpha F_{\rho \sigma}
+A_\rho F_{\sigma \alpha}
+A_\sigma F_{\alpha \rho}
\right)
\right]
\left(\partial_\beta A_\mu+2 F_{\beta\mu}\right)
\\ \nonumber
&=\frac{1}{6}
\partial_\rho
\left[
\theta^{\alpha\beta}\theta^{\sigma\rho} 
A_\sigma A_\alpha \left(\partial_\beta A_\mu+2 F_{\beta\mu}\right)\right]
-\frac{1}{12}\theta^{\alpha\beta}\theta^{\sigma\rho} 
\left(
A_\alpha F_{\rho \sigma}
+A_\rho F_{\sigma \alpha}
+A_\sigma F_{\alpha \rho}
\right)
\left(\partial_\beta A_\mu+2 F_{\beta\mu}\right)
\end{align}
Following the derivation of $A_\mu^{(1)}$ in Eqs.~\eqref{eq:A1} and~\eqref{eq:A1_D+CS}, we perform an integration by parts to the second term with $\partial_\beta A_\mu$. This gives
\begin{align}
\nonumber
A_\mu^{(2,2)}
=&~\frac{1}{6}
\partial_\rho
\left[
\theta^{\alpha\beta}\theta^{\sigma\rho} 
A_\sigma A_\alpha \left(\partial_\beta A_\mu+2 F_{\beta\mu}\right)\right]
-\frac{1}{12}\partial_\beta
\left[\theta^{\alpha\beta}\theta^{\sigma\rho} 
A_\mu\left(
A_\alpha F_{\rho \sigma}
+A_\rho F_{\sigma \alpha}
+A_\sigma F_{\alpha \rho}
\right)\right]
\\ \nonumber
&+\frac{1}{12}
\theta^{\alpha\beta}\theta^{\sigma\rho} 
A_\mu\left(
\partial_\beta A_\alpha F_{\rho \sigma}
+\partial_\beta A_\rho F_{\sigma \alpha}
+\partial_\beta A_\sigma F_{\alpha \rho}
\right)
-\frac{1}{6}
\theta^{\alpha\beta}\theta^{\sigma\rho} 
\left(
A_\alpha F_{\rho \sigma}
+A_\rho F_{\sigma \alpha}
+A_\sigma F_{\alpha \rho}
\right)F_{\beta\mu}
\\
&+\frac{1}{12}
\theta^{\alpha\beta}\theta^{\sigma\rho} A_\mu
\left(
A_\alpha \partial_\beta F_{\rho \sigma}
+A_\rho \partial_\beta F_{\sigma \alpha}
+A_\sigma  \partial_\beta F_{\alpha \rho}
\right)
\end{align}
The last term vanishes due to the antisymmetry of $\theta$ and $F$. The first line is clearly a sum of two boundary terms which can be further simplified. Using Eq.~\eqref{eq:identity1} and~\eqref{eq:identity2}, we have
\begin{align}
\label{eq:2ndOrder-CSform}
\nonumber
A_\mu^{(2,2)}
=&~\frac{1}{6}
\theta^{\alpha\beta}\theta^{\sigma\rho} 
\partial_\rho
\left[
A_\sigma A_\alpha \left(\partial_\beta A_\mu+2 F_{\beta\mu}\right)
-A_\mu\left(A_\sigma\partial_\beta A_\alpha + A_\beta F_{\alpha\sigma}\right)
\right]
\\ 
&+\frac{\hbar^2 }{6}\epsilon^{\alpha\beta\sigma\rho}
\left[A_\mu\partial_\alpha A_\beta\partial_\sigma A_\rho
+2A_\alpha\partial_\beta A_\sigma \partial_\rho A_\mu
-2A_\alpha \partial_\mu A_\beta \partial_\sigma A_\rho
\right]
\end{align}
In 5D, the expression can be written as a sum of derivative term and a CS 5-form:
\begin{align} \label{eq:5D-A2}
\nonumber
A_\mu^{(2,2)}
=&~\frac{1}{6}
\theta^{\alpha\beta}\theta^{\sigma\rho} 
\partial_\rho
\left[
A_\sigma A_\alpha \left(\partial_\beta A_\mu+2 F_{\beta\mu}\right)
-A_\mu\left(A_\sigma\partial_\beta A_\alpha + A_\beta F_{\alpha\sigma}\right)
\right]
\\ 
&+\frac{\hbar^2}{6}\delta_{\mu,0}
\epsilon^{\alpha\beta\sigma\rho\nu}
A_\alpha \partial_\beta A_\sigma \partial_\rho A_\nu 
\end{align}
Similar to $A_\mu^{(1)}=A_\mu^{(1,1)}$, the CS term vanishes if $\mu\neq 0$. When $d<5$, the CS term vanishes for all $\mu$. This completes our discussion of the second-order SW map.

\section{Chern-Simons term in the third order SW map}
\label{sec:3rdOrder}

In this Appendix, we show that the diagonal term in the third-order SW map for the gauge potential, $A_\mu^{(3,3)}$ in noncommutative phase space can be written as a sum of a total derivative and a term of the Chern-Simons form. For the case of 6+1-D phase-space, we obtain the topological CS term. 
From the recursive relation for diagonal terms \cite{Fidanza}, we have
\begin{eqnarray}
A_\gamma^{(3,3)}
=-\frac{1}{4}
\theta^{\alpha\beta}A_\alpha^{(2,2)}\left(\partial_\beta A_\gamma + 3F_{\beta\gamma}\right)
\end{eqnarray}

For the second-order term, $A_\alpha^{(2,2)}$, we use the form derived in Appendix \ref{app:2nd-SW} which makes the (4+1-D) CS term more explicit (Eq.~\eqref{eq:2ndOrder-CSform}),
\begin{align}
\nonumber
A_\alpha^{(2,2)}
=&~\frac{1}{6}
\theta^{\mu\nu}\theta^{\sigma\rho} 
\partial_\rho
\left[
A_\sigma A_\mu \left(\partial_\nu A_\alpha +2 F_{\nu\alpha}\right)
-A_\alpha\left(A_\sigma\partial_\nu A_\mu + A_\nu F_{\mu\sigma}\right)
\right]
\\ 
&+\frac{\hbar^2 }{6}\epsilon^{\mu\nu\sigma\rho}
\left[A_\alpha\partial_\mu A_\nu\partial_\sigma A_\rho
+2A_\mu\partial_\nu A_\sigma \partial_\rho A_\alpha
-2A_\mu \partial_\alpha A_\nu \partial_\sigma A_\rho
\right]
\end{align}
with the understanding that the $\epsilon^{\mu\nu\sigma\rho}$ antisymmetrizes nonzero indices in groups of four. For example, in 6+1-D, it will antisymmetrize the three groups
\begin{eqnarray}
    (\mu, \nu, \sigma, \rho) &\in (1,2,3,4),&\text{or}\\
                            &\in (1,2,5,6),&\text{or}\\ 
                            &\in (3,4,5,6). 
\end{eqnarray}

We can rewrite the CS form in $A_\alpha^{(2,2)}$ in terms of fields $F_{\mu\nu}$,
\begin{align}
\label{eq:secondOrderGauge}
\nonumber
    A_\alpha^{(2,2)}
    =&~\frac{1}{6}
    \theta^{\mu\nu}\theta^{\sigma\rho} 
    \partial_\rho
    \left[
    A_\sigma A_\mu \left(\partial_\nu A_\alpha +2 F_{\nu\alpha}\right)
    -A_\alpha\left(A_\sigma\partial_\nu A_\mu + A_\nu F_{\mu\sigma}\right)
    \right]
    \\ 
    \nonumber
    &+\frac{\hbar^2 }{6}\epsilon^{\mu\nu\sigma\rho}
    \left[\frac{1}{4}A_\alpha F_{\mu\nu}F_{\sigma\rho}
    +A_\mu F_{\nu\sigma}\partial_\rho A_\alpha
    -A_\mu  F_{\sigma\rho}\partial_\alpha A_\nu
    \right]  \\
    \nonumber
    =&~\frac{1}{6}
    \theta^{\mu\nu}\theta^{\sigma\rho} 
    \partial_\rho
    \left[
    A_\sigma A_\mu \left(\partial_\nu A_\alpha +2 F_{\nu\alpha}\right)
    -A_\alpha\left(A_\sigma\partial_\nu A_\mu + A_\nu F_{\mu\sigma}\right)
    \right]
    \\ 
    &+\frac{\hbar^2 }{24}\epsilon^{\mu\nu\sigma\rho}
    \left[A_\alpha F_{\mu\nu}F_{\sigma\rho}
    +4A_\mu F_{\nu\sigma}F_{\rho\alpha}
    \right]   
\end{align}
In the last equality, we have used the identity
\begin{equation}
    \epsilon^{\mu\nu\sigma\rho}A_\mu F_{\sigma\rho}\partial_\alpha A_\nu = \epsilon^{\mu\nu\sigma\rho}A_\mu F_{\nu\sigma}\partial_\alpha A_\rho.
\end{equation}

When applying the recursive formula, we need to keep in mind that the derivatives that are not part of gauge-invariant quantities like $F_{\mu\nu}$ act as operators from the left. 

\begin{align}
    \nonumber
    A_\gamma^{(3,3)} &= -\frac{1}{24}\theta^{\alpha\beta}\theta^{\mu\nu}\theta^{\sigma\rho}\left(
    \vec{\partial}_\rho
    \left[
    A_\sigma A_\mu \left(\vec{\partial}_\nu A_\alpha +2 F_{\nu\alpha}\right)
    -A_\alpha\left(A_\sigma\vec{\partial}_\nu A_\mu + A_\nu F_{\mu\sigma}\right)
    \right](\partial_\beta A_\gamma + 3F_{\beta\gamma})\right) \\ 
    & -\frac{\hbar^2 }{96}\theta^{\alpha\beta}\epsilon^{\mu\nu\sigma\rho}
    \left[A_\alpha F_{\mu\nu}F_{\sigma\rho}
    +4A_\mu F_{\nu\sigma}F_{\rho\alpha}
    \right] (\partial_\beta A_\gamma + 3F_{\beta\gamma}).
\end{align}

 We know that every term generated by the first term in Eq. \eqref{eq:secondOrderGauge} in the above equation can be written as a total derivative, so we drop those terms temporarily for the sake of simplicity. In our search for the CS terms we will continuously drop the boundary contributions going ahead. We will address the boundary terms after deriving the CS form. Let us define for convenience, 
 \begin{equation}
     \mathcal{A}_\gamma^{(3,3)} \equiv A_{\gamma}^{(3,3)} - \left(\mathrm{Bdy\ terms}\right).
 \end{equation}
We have, 
\begin{equation}
    \mathcal{A}_\gamma^{(3,3)} = -\frac{\hbar^2}{96}\theta^{\alpha\beta}\epsilon^{\mu\nu\sigma\rho}\left[A_\alpha F_{\mu\nu}F_{\sigma\rho}
    +4A_\mu F_{\nu\sigma}F_{\rho\alpha}
    \right] (\partial_\beta A_\gamma + 3F_{\beta\gamma}).
\end{equation}
Similar to the derivation of the CS term in the first and second-order SW maps, we integrate by parts the term with $\partial_\beta A_\gamma$. Up to the boundary terms, we get

\begin{align}
\begin{split}
\label{eq:thirdOrderTerms}
    \mathcal{A}_\gamma^{(3,3)} &= \frac{\hbar^2}{96}\theta^{\alpha\beta}\epsilon^{\mu\nu\sigma\rho}A_\gamma\left(\partial_\beta A_\alpha F_{\mu\nu}F_{\sigma\rho} + 4\partial_\beta A_\mu F_{\nu\sigma}F_{\rho\alpha}\right) - \frac{\hbar^2}{32}\theta^{\alpha\beta}\epsilon^{\mu\nu\sigma\rho}\left(A_\alpha F_{\mu\nu}F_{\sigma\rho}
    +4A_\mu F_{\nu\sigma}F_{\rho\alpha}\right)F_{\beta\gamma}\\
    &+\frac{\hbar^2}{96}\theta^{\alpha\beta}\epsilon^{\mu\nu\sigma\rho}\left(A_\alpha\partial_\beta(F_{\mu\nu}F_{\sigma\rho}) + 4A_\mu\partial_\beta(F_{\nu\sigma}F_{\rho\alpha})\right)
\end{split}
\end{align}

At this point, we verify a few identities explicitly for the 6-D phase space using \texttt{Mathematica} \cite{GithubRepo}. In the above equation, the last term sums to zero because of the anti-symmetry of $\theta^{\alpha\beta}$, $\epsilon^{\mu\nu\sigma\rho}$, and the field tensors. Thus we have,
\begin{equation}  
\label{eq:Identity1}
\theta^{\alpha\beta}\epsilon^{\mu\nu\sigma\rho}\left(A_\alpha\partial_\beta(F_{\mu\nu}F_{\sigma\rho}) + 4A_\mu\partial_\beta(F_{\nu\sigma}F_{\rho\alpha})\right) = 0
\end{equation}
when the Greek indices take nonzero values $(1,\ldots, 6)$. For the first term in Eq.~\eqref{eq:thirdOrderTerms}, we verified the identity
\begin{equation}
\label{eq:Identity2}
    \theta^{\alpha\beta}\epsilon^{\mu\nu\sigma\rho}\left(\partial_\beta A_\alpha F_{\mu\nu}F_{\sigma\rho} + 4\partial_\beta A_\mu F_{\nu\sigma}F_{\rho\alpha}\right) = -4\hbar\epsilon^{\alpha\beta\mu\nu\sigma\rho}\partial_\alpha A_\beta \partial_\mu A_\nu \partial_\sigma A_\rho
\end{equation}
where $\epsilon^{\alpha\beta\mu\nu\sigma\rho}$ is the fully anti-symmetric tensor in six indices. Similarly, the second term in Eq.~\eqref{eq:thirdOrderTerms} can be simplified as
\begin{equation}
\label{eq:Identity3}
    \theta^{\alpha\beta}\epsilon^{\mu\nu\sigma\rho}\left(A_\alpha F_{\mu\nu}F_{\sigma\rho}
    +4A_\mu F_{\nu\sigma}F_{\rho\alpha}\right)F_{\beta\gamma} = 4\hbar\epsilon^{\alpha\beta\mu\nu\sigma\rho}A_\alpha \partial_\beta A_\mu\partial_\nu A_\sigma(\partial_\rho A_\gamma - \partial_\gamma A_\rho)
\end{equation}
Now we can see the 7D CS term taking form. Using all three of the above identities together, we get
\begin{equation}
    \mathcal{A}_\gamma^{(3,3)} = -\frac{\hbar^3}{24}\epsilon^{\alpha\beta\mu\nu\sigma\rho}\left(A_\gamma\partial_\alpha A_\beta \partial_\mu A_\nu \partial_\sigma A_\rho + 3A_\alpha \partial_\beta A_\mu\partial_\nu A_\sigma(\partial_\rho A_\gamma - \partial_\gamma A_\rho)\right)
\end{equation}
which is nothing but the 7D CS term when $\gamma = 0$:
\begin{equation}
    \mathcal{A}_\gamma^{(3,3)} = -\frac{\hbar^3}{24}\delta_{\gamma,0}\epsilon^{\alpha\beta\mu\nu\sigma\rho\delta}A_\alpha \partial_\beta A_\mu \partial_\nu A_\sigma \partial_\rho A_\delta.
\end{equation}
Thus, the diagonal term $A_\gamma^{(3,3)}$ in the third-order Seiberg-Witten map can be written as the sum of a topological CS term in 7D and total derivatives.
For the sake of completeness, let us evaluate the boundary terms explicitly:
\begin{align}
\begin{split}
    A_{\gamma}^{(3,3)} - \mathcal{A}_\gamma^{(3,3)} &= -\frac{1}{24}\theta^{\alpha\beta}\theta^{\mu\nu}\theta^{\sigma\rho}\left(\partial_\rho\left[A_\sigma A_\mu\partial_\nu (A_\alpha \partial_\beta A_\gamma + 3A_\alpha F_{\beta\gamma}) + 2A_\sigma A_\mu F_{\nu\alpha}(\partial_\beta A_\gamma + 3F_{\beta\gamma})\right]\right)\\
    &+\frac{1}{24}\theta^{\alpha\beta}\theta^{\mu\nu}\theta^{\sigma\rho}\left(\partial_\rho\left[A_\alpha A_\sigma \partial_\nu(A_\mu\partial_\beta A_\gamma + 3F_{\beta\gamma}) + A_\alpha A_\nu F_{\mu\sigma}(\partial_\beta A_\gamma + 3 F_{\beta\gamma})\right]\right)\\
    &-\frac{\hbar^2}{96}\theta^{\alpha\beta}\epsilon^{\mu\nu\sigma\rho}\partial_\beta\left[A_\alpha A_{\gamma} F_{\mu\nu}F_{\sigma\rho} + 4A_\mu A_{\gamma}F_{\nu\sigma}F_{\rho\alpha}\right]
\end{split}
\end{align}

At the third-order in $\theta$, $A_\gamma^{(3,3)}$ is not the only term in the SW map, 
\begin{equation}
    A_\gamma^{(3)} = A_{\gamma}^{(3,3)} + A_\gamma^{(1,3)}.
\end{equation}
There is another term which is order 1 in $A$, given by \cite{Fidanza}
\begin{equation}
    A_\gamma^{(1,3)} = \frac{1}{48}\theta^{\alpha\beta}\theta^{\mu\nu}\theta^{\sigma\rho}(\partial_\alpha \partial_\mu A_\sigma)\left(\partial_\beta\partial_\nu(\partial_\rho A_\gamma + F_{\rho\gamma})\right)
\end{equation}
Integrating by parts, we see that $A_\gamma^{(1,3)}$ is infact a boundary term,
\begin{equation}
    A_\gamma^{(1,3)} = \frac{1}{48}\theta^{\alpha\beta}\theta^{\mu\nu}\theta^{\sigma\rho}\partial_\beta\bigg[(\partial_\alpha \partial_\mu A_\sigma)\partial_\nu(\partial_\rho A_\gamma + F_{\rho\gamma})\bigg].
\end{equation}

Thus, we arrive at the conclusion that the third-order SW map expansion for the $U(1)$ gauge field $A_0$ can be written as the sum of a Chern-Simons term in 7D and total derivatives,
\begin{equation}
    A_\gamma^{(3)} = -\frac{\hbar^3}{24}\delta_{\gamma,0}\epsilon^{\alpha\beta\mu\nu\sigma\rho\delta}A_\alpha \partial_\beta A_\mu \partial_\nu A_\sigma \partial_\rho A_\delta + (\mathrm{Bdy\ terms}).
\end{equation}

\section{CS forms in higher-order terms in the SW map: an incomplete proof}
\label{app:higherOrders}

In the main text and Appendices~\ref{app:2nd-SW} \&~\ref{sec:3rdOrder}, we have derived the CS terms for the 3,5, and 7 dimensional (phase) spacetime based on the first three orders of the Seiberg-Witten map. A natural follow-up question is: Can similar Chern-Simons terms in $2p+1$ dimensions be derived from the $p$-th order SW map? It is tempting to think that a similar procedure as in Appendix~\ref{sec:3rdOrder} can be generalized to arbitrary dimensions. In this section, we sketch an incomplete proof of this statement by induction.

Before we proceed, it is important to note that the derivation of the CS terms in the SW map at the second order and higher orders is only discussed here in the context where the noncommutativity matrix $\theta^{\alpha\beta}$ takes the symplectic form. That is, 
\begin{equation}
    \theta^{2i-1, 2i} = \hbar, \hspace{0.5in} i\in (1,2,\ldots d).
\end{equation}

Now, we begin the proof by induction. Assume that $A_\gamma^{(p-1,p-1)}$ can be written as a CS form in $2(p-1)+1$ dimensions, 
\begin{equation}
    A_\gamma^{(p-1,p-1)} = (-\hbar)^{d-1}A_{[\gamma}\partial_{\mu_1}A_{\mu_2}\partial_{\mu_3}A_{\mu_4}\ldots \partial_{\mu_{2p-3}}A_{\mu_{2p-2]}} + (\mathrm{Bdy\ terms})
\end{equation}
where the square brackets denote anti-symmetrization of the indices. Given the above form and the recursive relation
\begin{equation}
    A_\gamma^{(p,p)} = -\frac{1}{p+1}\theta^{\alpha\beta}A_\alpha^{(p-1,p-1)}(\partial_\beta A_\gamma + p\cdot F_{\beta\gamma}),
\end{equation}
we need to prove that the resulting diagonal term is also of a CS form. The edge case is proved in Appendix~\ref{sec:3rdOrder}, where we showed that the CS form in the third-order SW map can be derived from the second-order terms and the recursive relation. 

The boundary terms from $A_\gamma^{(p-1, p-1)}$ become boundary terms in $A_\gamma^{(p,p)}$ because derivatives enter as operators in the recursive relation. We ignore the boundary terms. Next, we must integrate by parts the term with $\partial_\beta A_\gamma$. This is possible if the higher-dimensional equivalents of the identities Eqs. \eqref{eq:Identity1}, \eqref{eq:Identity2}, \eqref{eq:Identity3}  are proven. Here, we state them as conjectures.

\emph{Conjecture C.1}:
We conjecture that the terms with derivatives of field tensor that are generated due to the integration by parts of $\partial_\beta A_\gamma$ sum up to zero in arbitrary dimensions. Formally, 
\begin{equation}
    \theta^{\alpha\beta}\epsilon^{\mu_1\ldots \mu_{2p-2}}\left(A_\alpha\partial_\beta\left(\prod_{i = 1}^{p-1}F_{\mu_{2i-1},\mu_{2i}}\right) + 2(p-1)A_{\mu_1}\partial_\beta\left(F_{\mu_{2p-2},\alpha}\prod_{i=2}^{p-2}F_{\mu_{2i},\mu_{2i+1}}\right)\right) = 0
\end{equation}
where the anti-symmetric tensor $\epsilon$ is understood to anti-symmetrize a set of $2(p-1)$ indices at a time (there will be $p$ such combinations). 

\emph{Conjecture C.2}:
Next, we conjecture the analogue of Eq.~\eqref{eq:Identity2},
\begin{equation}
    \theta^{\alpha\beta}(\partial_\beta A_{[\gamma})\partial_{\mu_1}A_{\mu_2}\ldots \partial_{\mu_{2p-3}}A_{\mu_{2p-2}]} = -\frac{\hbar}{(2p-1)!} \epsilon^{\mu_1\ldots \mu_{2p}}\prod_{i=1}^{2p-1}\partial_{\mu_i}A_{\mu_{i+1}}.
\end{equation}

\emph{Conjecture C.3}:
Finally, we have the higher-dimensional analogue of Eq.~\eqref{eq:Identity3}
\begin{multline}
    \theta^{\alpha\beta}\epsilon^{\mu_1\ldots \mu_{2p-2}}\left( A_\alpha\prod_{i = 1}^{p-1}F_{\mu_{2i-1},\mu_{2i}} + (p-1)A_{\mu_1}F_{\mu_{2p-2},\alpha}\prod_{i=2}^{p-2}F_{\mu_{2i}, \mu_{2i+1}}\right)F_{\beta\gamma} \\= 2^{p-1}\hbar \epsilon^{\mu_1\ldots \mu_{2p}}A_\alpha\partial_\beta A_{\mu_1}(\partial_{\mu_{2p-2}}A_\gamma - \partial_\gamma A_{\mu_{2p-2}})\prod_{i=2}^{p-2}\partial_{\mu_{2i}}A_{\mu_{2i+1}}
\end{multline}
Using these identities in the recursive formula, it is easy to see that

\begin{equation}
    A_\gamma^{(p,p)} = -\hbar^p A_{[\gamma}\partial_{\mu_1}A_{\mu_2}\ldots \partial_{\mu_{2p-1}}A_{\mu_{2p}]} + (\mathrm{Bdy\ terms})
\end{equation}
When the indices take values $\mu_i \in (1,2,3,\ldots 2p)$. Due to the anti-symmetry of $\gamma$ with respect to $\mu_i$, the CS term only contributes when $\gamma = 0$. Thus, we can write

\begin{equation}
    A_\gamma^{(p, p)} = \frac{(-\hbar)^{p}}{(p+1)!}\delta_{\gamma,0}\epsilon^{\mu_1, \ldots \mu_{2p+1}}A_{\mu_1}\partial_{\mu_2}A_{\mu_3}\ldots \partial_{\mu_{2p}}A_{\mu_{2p+1}} + (\mathrm{Bdy\ terms})
\end{equation}
where $\mu_i$ now take values $(0,1,2,\ldots, 2p)$. Thus, starting from the form of $A_0^{(p-1,p-1)}$, we have shown that the $p$-th order diagonal term $A_0^{(p,p)}$ can be rewritten as a topological CS-term in $2p + 1$-dimensional spacetime, upto total derivatives. The proof is strictly incomplete, it relies strongly on the conjectures we made. Nevertheless, the recursive nature of the procedure hints that there might be a deep mathematical reason for the appearance of the CS forms in the SW map at all orders. In particular, Ref.~\cite{Picariello2002} pointed out that the SW map of a noncommutative Abelian gauge field can be formulated in the language of Becchi-Rouet-Stora-Tyutin (BRST) cohomology. In this formulation, the topologically nontrivial CS terms derived from the Seiberg-Witten map can be understood as different nontrivial elements in the BRST cohomology groups. In our present work, there is an additional symplectic structure for the noncommutative phase space. We are not aware of any mathematical proof for the emergence of CS terms in every order of the Seiberg-Witten map, and how it can be derived rigorously through BRST cohomology. We will leave this as a speculation and an open question to future work.

\twocolumngrid

\end{document}